\documentclass[a4paper,11pt]{article}

\usepackage{amsmath} % assumes amsmath package installed
\usepackage{amssymb}  % assumes amsmath package installed
 \usepackage[english]{babel}
 \usepackage{dsfont}
\usepackage{amsfonts,amsmath,amsthm,graphicx,a4wide}   
\usepackage{braket}
\usepackage{xcolor}
\usepackage{bbold}
\usepackage{graphicx}
\usepackage{braket}
\usepackage{color}
\usepackage{graphicx}

\newtheorem{thm}{Theorem}[section]

\newtheorem{rem}[thm]{Remark}

\allowdisplaybreaks
\usepackage{lipsum}
\newcommand{\tr}{\operatorname{Tr}}
\newcommand{\Tr}{\operatorname{Tr}}
\renewcommand{\H}{\mathcal H}
\newcommand{\F}{\mathcal F}
\newcommand{\K}{\boldsymbol K}
\newcommand{\E}{\mathbb E}
\renewcommand{\P}{\mathbb P}
\newcommand{\Q}{\mathbb Q}
\newcommand{\1}{\mathbb 1}
\renewcommand{\o}{\left(}
\renewcommand{\c}{\right)}

\newcommand{\p}{p^*}
\newcommand{\J}{\mathcal J}

\newcommand{\Y}{\mathcal Y}
\newcommand{\I}{\mathcal I}
\renewcommand{\p}{p^*}
\newcommand{\V}{\mathcal V}
\renewcommand{\F}{\mathcal F}

\title{\LARGE \bf
Parameter Estimation for Quantum Trajectories: Convergence Result}

\author{Ma\"{e}l Bompais \and Nina H. Amini \thanks{Ma\"el Bompais and Nina H. Amini are with Laboratoire des signaux et syst\`{e}mes (L2S), CNRS-CentraleSup\'{e}lec-Universit\'{e} Paris-Sud, Universit\'{e} Paris-Saclay, 3, rue Joliot Curie, 91190 Gif-sur-Yvette, France
  ({first name.last name@centralesupelec.fr}.)} \and Cl\'ement Pellegrini\thanks{Cl\'ement Pellegrini is with Institut de Math\'ematiques, IMT, Université de Toulouse (UMR 5219), 31062 Toulouse, Cedex 9, France (clement.pellegrini@math.univ-toulouse.fr).}
}

\date{}
\begin{document}

\maketitle
\thispagestyle{empty}
\pagestyle{empty}

%%%%%%%%%%%%%%%%%%%%%%%%%%%%%%%%%%%%%%%%%%%%%%%%%%%%%%%%%%%%%%%%%%%%%%%%%%%%%%%%
\begin{abstract}
A quantum trajectory describes the evolution of a quantum system undergoing indirect measurement. In the discrete-time setting, the state of the system is updated by applying Kraus operators according to the measurement results. From an experimental perspective, these Kraus operators can depend on unknown physical parameters $p$. An interesting and powerful method has been proposed in \cite{six2015parameter} to estimate a parameter in a finite set; however, complete results of convergence were lacking. This article fills this gap by rigorously showing the consistency of the method, whereas there was only numerical evidence so far. When the parameter belongs to a continuous set, we propose an algorithm to approach its value and show simulation results.
\end{abstract}

%%%%%%%%%%%%%%%%%%%%%%%%%%%%%%%%%%%%%%%%%%%%%%%%%%%%%%%%%%%%%%%%%%%%%%%%%%%%%%%%
\section{Introduction}
Parameter estimation is a fundamental subject in Statistics. In the context of quantum trajectories, it can be formulated as parameter estimation for a particular hidden Markov chain. The system of interest is not directly observed and a measurement apparatus captures only partial information \cite{belavkin1983,belavkin1989nondemolition,belavkin1995quantum,belavkin1992quantum,barchielli2009quantum}.  

The theory of parameter estimation is well studied for hidden Markov models  \cite{rabiner1989tutorial,leroux1992maximum}. For instance, many papers established the consistency and asymptotic normality of the maximum likelihood estimator \cite{bickel1996inference,bickel1998asymptotic,fredkin1992maximum}. Parameter estimation is a rapidly developing field in the context of quantum dynamical systems with central application in quantum metrology  \cite{giovannetti2011advances,demkowicz2012elusive,vidrighin2014joint}. For example, achieving the Heisenberg limit is a key challenge \cite{zhou2018achieving,giovannetti2004quantum}. %and key impact in real experiments \cite{vidrighin2014joint}. %The estimation problem can be stated for open quantum systems, that is systems which are interaction with environments \cite{}. 

Theoretical investigations regarding estimation of an unknown physical parameter through continuous measurement have been initiated by Mabuchi for identifying an effective Hamiltonian in the context of cavity quantum electrodynamics \cite{mabuchi1996dynamical}. Another series of papers has further developed identification of dynamical quantum systems \cite{burgarth2012quantum,zhang2014quantum,wang2017quantum,nurdin2020data,heatwolequantum,gough2009series,guctua2015system,guta2017information}.  

Parameter estimation has been developed in various directions in quantum mechanics. %Concerning state tomography from the measurements records, we refer to \cite{gambetta2001state,chase2009single,six2015parameter,metillon2019benchmarking,burgarth2015quantum,six2016quantum}.]  
In the context of quantum non-demolition measurements, in \cite{benoist2018quantum}, the authors show the asymptotic normality and consistency of the maximum likelihood estimator. In the framework of quantum statistics, local asymptotic normality has been investigated in \cite{guctua2007local}. Regarding quantum Fisher information and Cram\'er Rao bound notions in the quantum dynamical systems framework, we refer to \cite{holevo2003statistical}. %The estimation might be defined in the formalism of input-output theory, where the Hamiltonian and coupling operator depend on unknown parameters. By measuring an output of the system, unknown physical parameters of the dynamical systems can be estimated.  
%The citation further extended/extends quantum fisher information.....

This paper focuses on the estimation problem for discrete-time quantum trajectories, where unknown parameters appear in the Kraus operators. We consider the same problem which was previously introduced in \cite{gambetta2001state},\cite{chase2009single},\cite{six2015parameter} in the aim of parameter discrimination. We apply the same formalism considered in these papers, meaning that we define an extended quantum trajectory on a larger Hilbert space, and we suppose that the initial state is unknown. In particular, we make use of the block structure proposed in \cite{six2015parameter} and we work with a prior law on the set of parameters. In  \cite{six2015parameter}, the authors show that the posterior law on the true parameter multiplied by the fidelity between the true trajectory and the estimated one is a submartingale, i.e., this quantity is non-decreasing in average. We prove that this posterior law on the true parameter converges to one.

%DEFINE PARAMETER DISCRIMINATION, HOW IT IS DIFFERENT FROM PARAMETER ESTIMATION [FROM CALVIN]
Our approach differs from the previous ones by employing a weaker condition than the asymptotic stability of quantum trajectories. For example, the asymptotic stability has been originally applied in \cite{chase2009single} in the aim of parameter discrimination. In  \cite{six2015parameter}, the parameter discrimination is related in some ways to the study of asymptotic stability as the authors work with fidelity. They show that such discrimination can be achieved in a non-decreasing manner in average; however, the complete convergence proof was missing. 
%[paragraph break here...?]
Here using an \textit{identifiability} condition, we prove that the complete convergence towards the true parameter can be achieved for the case where the unknown parameter belongs to a finite set. Roughly speaking, this condition means that each parameter gives rise to an observation process whose law is different for two distinct parameters. In particular, we make the condition precise in terms of invariant states of quantum channels \cite{wolf2012quantum}. %Quantum channels refer to conditional expectation of quantum trajectories knowing the previous state, see, \cite{wolf2012quantum}. [PLURAL OR SINGULAR CONDITION? CHECK] 
 In addition, we provide the convergence speed of this selection of true parameter. Our contribution closes the underlined issue in \cite{six2015parameter} since we show rigorously the result of convergence. In particular, our paper confirms that the technique developed in \cite{six2015parameter} is efficient. We go further by exhibiting  an exponential speed of convergence. To this end, we adapt ergodic theory tools as initiated in \cite{benoist2018entropy}.

The paper is structured as follows. Section \ref{sec:des} presents a model description of discrete-time quantum trajectories and preliminary tools. In Section \ref{sec:three}, we show the convergence towards the true parameter under the identifiability condition (Theorem \ref{thm:convergence}). Section \ref{sec:speed} gives the speed of convergence (Theorem \ref{thm:speed}). In Section \ref{sec:algo}, we discuss discrimination of parameters through a heuristic algorithm and simulations in the case where the unknown parameter can take values in an infinite set. Finally, Section \ref{sec:conclusion} concludes this work.

\section{Model description}
\label{sec:des}
This section presents dynamics of discrete-time quantum trajectories and states the estimation problem. 
\subsection{Quantum trajectory}
We consider the Hilbert space $\H=\mathbb C^d$ describing a finite-dimensional quantum system undergoing repeated indirect measurements. The state evolution of the system is given by the following Markov chain\begin{equation}
{\rho}_{n+1}=\frac{\boldsymbol{K}_{y_{n}}^{{}}\left({\rho}_{n}\right)}{\operatorname{Tr}\left(\boldsymbol{K}_{y_{n}}^{{}}\left({\rho}_{n}\right)\right)}.
\label{eq:true}
\end{equation}
Here
\begin{itemize}
\item $\rho_n$ represents quantum state at time $n$ and belongs to the set of density matrices $\mathcal S(\H):=\{\rho\in\mathbb C^{d\times d}|\,\rho=\rho^\dag,\, \rho \geq 0,\, \tr(\rho)=1\}.$
\item $y_n$ corresponds to the measurement result at step $n$ and takes values in a finite alphabet $\mathcal Y.$
\item The probability of $y_n=y$ is given by $\mathrm{Tr}(\boldsymbol{K}_{y}(\rho_n)).$
\item The $\boldsymbol{K}_{y}$ are the Kraus operators: for $y \in \Y$, $\boldsymbol{K}_{y}(\rho)=\sum_{\mu}V_{y,\mu}\rho V_{y,\mu}^\dag.$
\item  $V_{y,\mu} \in\mathbb C^{d\times d}$ and $\sum_{y,\mu}V_{y,\mu}^\dag V_{y,\mu}=\1,$ where $\1$ denotes the identity operator.
\end{itemize}
~~\\
The sequence of states $(\rho_n)_{n\geq 0}$ is called a quantum trajectory. The formulation with operators ${\K_y}$ takes into account possible measurement imperfections and/or partially read measurements. For perfect measurements, $\K_y(\rho)=V_y \rho V_y^\dag$ and for imperfect measurements, $\K_y(\rho)=\sum_{ \mu}\eta_{y,\mu}V_{\mu}\rho V_{\mu}^\dag$, where $\eta$ denotes the correlation matrix (see e.g. \cite{somaraju2012design}).\\
An important feature of a quantum trajectory is the associated quantum channel $\Phi=\sum_{y}\K_y$. It is a completely positive and trace-preserving map, characterizing the conditionnal expectation of $\rho_{n+1}$ knowing $\rho_n$:
$$\E[\rho_{n+1}|\rho_n]=\Phi(\rho_n).$$

Under experimental conditions, it might happen that the initial state $\rho_0$ of the physical system is unknown, hence $(\rho_n)$ cannot be directly computed. In this situation, we build an estimation $(\hat \rho_n)$ of the quantum trajectory by setting an arbitrary initial state $\hat \rho_0$ and make it evolve according to the measurement results coming from the system:
\begin{equation}
{\hat \rho}_{n+1}=\frac{\boldsymbol{K}_{y_{n}}^{{}}\left({\hat\rho}_{n}\right)}{\operatorname{Tr}\left(\boldsymbol{K}_{y_{n}}^{{}}\left({\hat\rho}_{n}\right)\right)},
\label{eq:estimate}
\end{equation}
where the $(y_n)$ involved correspond to trajectory \eqref{eq:true}. To prevent the denominator from vanishing, we usually require $\ker \hat \rho_0 \subset \ker \rho_0$. A natural choice for this purpose is to set $\hat \rho_0$ as the completely mixed state: $\hat \rho_0 = \frac{\1}{d}$.
The asymptotic behaviour of this estimated trajectory has been studied in \cite{amini2021asymptotic}; in particular the authors showed that it converges to the actual trajectory under a purification assumption.
\subsection{Problem setting} The goal of this paper is to consider the situation where the Kraus operators $\{\boldsymbol{K}_{y}\}$ depend on an unknown physical parameter $p$, and try to estimate it. We mark the dependence in $p$ as follows
\begin{equation}
    {\rho}_{n+1}=\frac{\boldsymbol{K}_{y_{n}}^{{p}}\left({\rho}_{n}\right)}{\operatorname{Tr}\left(\boldsymbol{K}_{y_{n}}^{{p}}\left({\rho}_{n}\right)\right)},
\quad
{\hat\rho}_{n+1}=\frac{\boldsymbol{K}_{y_{n}}^{{p}}\left({\hat\rho}_{n}\right)}{\operatorname{Tr}\left(\boldsymbol{K}_{y_{n}}^{{p}}\left({\hat\rho}_{n}\right)\right)}
\label{eq:p}
\end{equation}
The parameter $p$ might be scalar or vectorial; we denote by $\p$ its exact value. We consider that the initial state $\rho_0$ of the system is unknown, hence the only pieces of information available are the measurement records $(y_0,y_1,y_2,...)$ obtained over time: this is a typical situation of a hidden Markov model. The following section provides a sufficient condition allowing the determination of $\p$.
In order to formalize our results, we state the following preliminary tools. 
\subsection{Preliminary tools}
\paragraph{Probability space} Let put $\Omega=\mathcal{Y}^{\mathbb{N}}$. A state $\rho$ and a parameter p generate a probability measure $\P_{\rho}^p$ on the finite sequences of elements of $\Y$:
$$
\mathbb{P}_{\rho}^p\left(y_{0},\ldots,y_{n-1}\right)=\tr ( \K_{y_{n-1}}^p\circ\ldots\circ\K_{y_0}^p (\rho ) )
$$
with $\circ$ denoting the composition of operators.
 More precisely, if we define the cylinder set $C_{y_{0}, \ldots, y_{n-1}}=\left\{\omega \in \Omega \mid \omega_{0}=y_{0}, \ldots, \omega_{n-1}=y_{n-1}\right\}$ of size $n$, the above expression defines a probability measure on the $\sigma$-algebra  $\mathcal{F}_{n}$ generated by all these cylinder sets. It can be extended to a probability measure, still denoted by $\P_{\rho}^p$, on the whole cylinder algebra $\mathcal F$ generated by all the $\mathcal{F}_{n}^{\prime} s$ (using the consistency Kolmogorov extension Theorem). This probability measure allows to consider asymptotic events for a quantum trajectory with Kraus operators $\{\K_y^p\}$ taking $\rho$ as an initial state. %Note that $\left(\Omega,\left(\mathcal{F}_{n}\right), \mathbb{P_{\rho}^p}\right)$ is a filtered probability space.
 \\
\paragraph{Minimal subspaces}
An invariant state $\rho_{inv}$ is defined as a fixed point of the quantum channel:
$\Phi(\rho_{inv})=\rho_{inv}.$
Throughout this paper, we consider that every quantum channel is \textit{faithful}, that is there exists a full-rank invariant state, or, equivalently, no transient part (see \cite{baumgartner2012structures}). In this situation, the Hilbert space can be decomposed in a unique fashion $\H=\underset{i=1}{\overset{m}{\bigoplus}} \V_i$, where $\V_i$ is stable by $\Phi$ and is the support of a unique \textit{minimal} invariant state $\rho_{inv}^i$. By minimal, we mean there does not exist any other invariant state with support strictly included in $\V_i$.
In addition, we define $M_i$ the orthogonal projector on $\V_i$. Finally, in order not to carry the heavy notation $\P_{\rho_{inv}^{p,i}}^p$, we denote by $\P_i^p$ the probability measure generated by the unique invariant state of the quantum channel $\Phi^p$ with support $\V_i^p$.
\\
\paragraph{A selection theorem and an ergodic theorem}

Here we recall our theorem in \cite{amini2021asymptotic} regarding the selection of a minimal subspace.
\begin{thm}[Amini et al., 2021]
Consider a quantum trajectory $(\rho_n)$ described by the Markov chain \eqref{eq:true}. Denote respectively by $\V_i$, $M_i$ and $\P_i$ the minimal subspaces, the corresponding orthogonal projectors, and the probability measures generated by the associated minimal invariant state, all labelled by $i\in \{1,...,m\}$.  Assume $\P_{i} \neq \P_{j}$ for $i\neq j$. Then there exists a random variable $\I$ valued in $\{1,...,m\}$ such that:
$$\lim\limits_{n\to \infty}\tr(M_{\I}\rho_n)=1 ~~\P_{\rho_0}-a.s.$$
with $\P_{\rho_0}(\I=i)=\tr(M_i\rho_0)$. Moreover, if $\hat\rho_0>0$,
$$\lim\limits_{n\to \infty}\tr(M_{\I}\hat\rho_n)=1~~\P_{\rho_0}-a.s.$$
\end{thm}
\medskip\begin{rem}
We will say that the trajectory $(\rho_n)$ \textit{selects} the minimal subspace $\V_i$ when $\lim\limits_{n\to \infty}\tr(M_{i}\rho_n)=1$.
\end{rem}
\medskip
\begin{rem}
The hypothesis $\P_{i} \neq \P_{j}$ for $i \neq j$ can be relaxed: if there exists $i_1\neq i_2$ in $\{1,...,m\}$ such that $\P_{{i_1}}=\P_{{i_2}}$, we may group them into an equivalence class, and the same result holds with the projectors on the direct sums of minimal subspaces belonging to the same equivalence class.
\end{rem}
\medskip
The following theorem recalls Kingman's subadditive ergodic theorem \cite{schurger1991almost}. This will be applied to derive the speed of convergence in Section \ref{sec:speed}. We introduce $\theta$ the shift operator:
$$\theta((\omega_0,\omega_1,\omega_2,...))=(\omega_1,\omega_2,\omega_3,...)$$
and denote by $\theta^n$ the composition $\theta \circ ...\circ \theta$ repeated $n$ times.\\
\begin{thm}[Kingman's subadditive ergodic theorem]
Let $(\Omega,\P,\theta)$ be an ergodic dynamical system. Let $(g_n)_{n\geq 0}$ be an almost subadditive sequence of integrable functions on $\Omega$, that is there exists some real number c such that
$$g_{n+m}(\omega) \leq g_n(\omega) + g_m(\theta^n(\omega))+c.$$
Then:
$$\frac{1}{n}g_n \xrightarrow[n\to \infty]{\P-a.s.,~L^1}\lim\limits_{n\to \infty}\frac{1}{n}\E[g_n].$$
\end{thm}
\medskip
\section{Parameter in a finite set}
\label{sec:three}
In this section, we assume that the parameter $\p$ takes one of the values in the finite set $\mathcal P =\{p_1,...,p_r\}$. We begin by recalling the structure of quantum filters designed for parameter discrimination, used for example in  \cite{gambetta2001state},\cite{chase2009single},\cite{six2015parameter}. The main idea consists in embedding the set of parameters $\mathcal P$ into a larger Hilbert space $\H_{\mathcal P} \otimes \H $, where $\H_{\mathcal P}$  is an abstract Hilbert space whose dimension is equal to the number of elements in  $\mathcal P$. In the sequel, for the sake of clarity, we will take $\mathcal P=\{a,b\}$ a set of two elements, leading to a tensorial space $\H_{a,b} \otimes \H  $ of dimension 2d. We will get convinced that every proof can be extended to any set $\mathcal P$ of finite cardinality.\\
On this tensorial Hilbert space $ \H_{a,b}\otimes \H $, we build an abstract block-diagonal quantum trajectory $(\Xi_n)$ evolving according to Kraus operators $\tilde K_y=diag(K_y^a,K_y^b)$, i.e.

$$\tilde{\boldsymbol{K}}_{y}(\Xi)=\sum_{\mu \in \Y}\left(\begin{array}{c|c}
V_{y,\mu}^{a} & 0 \\
\hline 0 & V_{y,\mu}^{b}
\end{array}\right) \Xi\left(\begin{array}{c|c}
V_{y,\mu}^{a} & 0 \\
\hline 0 & V_{y,\mu}^{b}
\end{array}\right)^{\dag}$$
and
\begin{equation}
    {\Xi}_{n+1}=\frac{\boldsymbol{\tilde K}_{y_{n}}^{{}}\left({\Xi}_{n}\right)}{\operatorname{Tr}\left(\boldsymbol{\tilde K}_{y_{n}}^{{}}\left({\Xi}_{n}\right)\right)}.
    \label{Xi_traj}
\end{equation}

Strictly speaking, the measurement results are emitted by the trajectory $(\rho_n)$
initialized with $\rho_0$ and with parameter $p=\p.$ However, they can be rigorously seen as coming from the trajectory $(\Xi_n)$
initialized with $\rho_0$ in the block of the true parameter -let us say for example $\p =a$-
$$\Xi_0=\left(\begin{array}{c|c}
\rho_0 & 0 \\
\hline 0 & 0
\end{array}\right).\\
$$
\medskip
More precisely, the probability measures induced on the set $\Omega$ by initial state $\Xi_0$ and Kraus operators $\{\tilde\K_y\}$ on one side and $\rho_0$ and $\{\K_y^{\p}\}$ on the other side are equal: $\P_{\Xi_0}^{(a,b)}=\P_{\rho_0}^{\p}.$ Hence, an estimated trajectory $(\hat\Xi_n)$ run with measurement records $(y_0,y_1,...)$ coming from the system (i.e., emitted according to $\P_{\rho_0}^{\p}$) can be seen as an estimation of the autonomous Markov chain $(\Xi_n)$. We set an estimated starting state $\hat \Xi_0$ spread out between the two blocks corresponding to a and b
$$\hat\Xi_0=\left(\begin{array}{c|c}
\pi_0^a\hat \rho_0 & 0 \\
\hline 0 & \pi_0^b\hat \rho_0
\end{array}\right)\\
$$
with $\hat \rho_0$ any full rank state on $\H$  and $(\pi^a_0,\pi^b_0)$ a \textit{prior law} on $\{a,b\}$: $(\pi^a_0,\pi^b_0) \in[0,1]^2$ and $\pi^a_0+\pi^b_0=1$. Then the time evolution may be written as
\begin{equation}
    \hat{\Xi}_{n}=\left(\begin{array}{c|c}
    \pi_{n}^a \hat{\rho}_{n}^{a} & 0 \\
    \hline 0 & \pi_{n}^b \hat{\rho}_{n}^{b}
    \end{array}\right)
    \label{Xi_traj_est}
\end{equation}\\
with a state $\hat{\rho}_{n}^{p}$ evolving according to operators $\{\K_y^{p}\}$ and $\pi_n^p=\tr (M^p \hat\Xi_n)$, where $M^p:=\sum_{i=1}^{m_p}M_i^p$ for $p=a,b$.
The coefficients $\pi_n^a$ and $\pi_n^b$ vary over time due to the normalization taking in account both blocks at each step.\\

We now state the identifiability assumption. Recall that the minimal subspaces $\V_i^p$ of a quantum channel $\Phi^p$ are labelled by $i \in \{1,...,m_p\}$, and that $\P_i^p$ denotes the probability measure generated by the unique invariant state with support $\V_i^p$.\\

\medskip
\textbf{Assumption (ID):}
For any $i \in \{1,...,m_a\}$ and $j \in \{1,...,m_b\}$, we assume
$$\P^a_{i} \neq \P^b_{j}.$$
Assumption (ID) may be reformulated as follows: for any minimal invariant state $\rho_{inv}^{a,i}$ of $\Phi^a$ and $\rho_{inv}^{b,j}$ of $\Phi^b$,
there exists an integer $l>0$  and a sequence $(y_1,\ldots,y_l) \in \Y^l$ such that
$$
 \tr ( \boldsymbol{K}^{\boldsymbol{a}}_{y_l} \circ ...\circ \boldsymbol{K}^{\boldsymbol{a}}_{y_1} ( \rho_{inv}^{a,i} ) ) \neq \tr ( \boldsymbol{K}^{\boldsymbol{b}}_{y_l} \circ ...\circ \boldsymbol{K}^{\boldsymbol{b}}_{y_1} ( \rho_{inv}^{b,j} ) ).
$$
%\medskip
Now we are ready to state our main theorem.
\medskip
\begin{thm}
Consider the trajectory $(\hat \Xi_n)$ initialized with $\pi^a_0>0$, $\pi^b_0>0$ and a full rank state $\hat \rho_0$. Denote by ${\p}\in \{a,b\}$ the true value of the parameter. Assume \textbf{(ID)} holds.
Then
$$\lim\limits_{n \to \infty}  \pi_{n}^{\p} = 1$$
and for $p\neq\p$:
$$\lim\limits_{n \to \infty}  \pi_{n}^{p} = 0$$
$\P^{\p}_{\rho_0}$ almost surely.
\label{thm:convergence}
\end{thm}
\medskip
\begin{proof}
Apply theorem 2.1 to the quantum trajectory $(\Xi_n)$. Because of its block-diagonal construction, the minimal subspaces are the gathering of the ${\V_i^a}^{\prime} s$ and the ${\V_i^b}^{\prime} s$. Let us say $\p=a$. Since $\tr(M_i^b \Xi_0)=0$ for all $i \in \{1,...,m_b\}$, the trajectory $(\Xi_n)$ almost surely never selects one of the ${\V_i^b}^{\prime} s$. Then there exists a random variable $\I$ valued in $\{1,...,m_a\}$ such that $\tr(M_{\I}^a \Xi_n)\xrightarrow[n\to\infty]{}1$ and, since $\hat\Xi_0>0$, $\tr(M_{\I}^a \hat\Xi_n)\xrightarrow[n\to\infty]{}1$ as well. As $\pi_n^a=\sum_{i=1}^{m_a}\tr(M_i^a\hat\Xi_n)$ and $\pi_n^b=\sum_{i=1}^{m_b}\tr(M_i^b\hat\Xi_n)$ and $\pi_n^a+\pi_n^b=1$, we have $\pi_n^a \xrightarrow[n\to\infty]{}1$ and $\pi_n^b \xrightarrow[n\to\infty]{}0$. Mutatis mutandis, when $\p=b$, then $\Xi_0=diag(0,\rho_0)$ and therefore $\pi_n^a \xrightarrow[n\to\infty]{}0$ and $\pi_n^b \xrightarrow[n\to\infty]{}1$.
\end{proof}
\bigskip
\begin{rem}
In \cite{six2015parameter}, the authors showed that $\pi_{n}^{\p} F(\rho_n,\hat \rho_n)$ is a sub-martingale, where $F$ denotes the fidelity (\cite{nielsen2002quantum}) between quantum states. In other words $\pi_{n}^{\p} F(\rho_n,\hat \rho_n)$ is non-decreasing in average. This means even if asymptotic stability is ensured, i.e., $\lim_{n\rightarrow \infty}F(\rho_n,\hat\rho_n)=1,$ we can only guaranty the increase in average of $\pi_n^{\p}.$ The above theorem shows $\pi_n^{\p}$ converges to one with a weaker condition (identifiability assumption) than the ones required for stability such as purification. 
\end{rem}
\section{Convergence speed}
\label{sec:speed}
The speed of selection of the true parameter can be captured by the limit of the quantity
$$
\frac{1}{n} \log \left(\frac{\pi_n^{p}}{\pi_{n}^{\p}}\right), ~p \neq  \p
$$
Roughly speaking, assume that there exists some constant $C>0$ such that
$$
\lim\limits_{n \to \infty}\frac{1}{n} \log \left(\frac{\pi_{n}^p}{\pi_{n}^{\p}}\right) =-C
$$
Then for large $n$
$$
\pi_{n}^p\simeq\pi_{n}^{\p} e^{-n C}
$$\\
Let us denote by $\P|_{\F_n}$ the probability $\P$ restricted to the cylinder $\sigma$-algebra $\F_n$. We introduce the quantities $S(\P):=\lim\limits_{n \to \infty} \frac{1}{n} H(\left. \P \right|_{\F_n})$, where $H$ is the \textit{Shannon entropy} and, for two probabilities $\P$ and $\Q$, $S(\P,\Q):=\lim\limits_{n \to \infty} \frac{1}{n}D \o \left. \P\right|_{\F_n}\left|\right|\left. \Q\right|_{\F_n} \c$, where $D$ is the \textit{Kullback-Leibler divergence} also called \textit{relative entropy}. 
\medskip
\begin{thm} For $p\neq {\p}$, we have
$$ \lim\limits_{n \to \infty}\frac{1}{n} \log \o \frac{ \pi_{n}^p}{ \pi_{n}^{\p}} \c \leq - ~\underset{i,j}{min}~S(\P_{i}^{\p},\P_{j}^{p})$$
$\P_{\rho_0}^{\p}$ almost surely, where the minimum is taken over $i \in \{1,...,m_p\}$ and $j\in\{1,...,m_{\p}\}$ and the right-hand side is a strictly negative constant under (ID).
\label{thm:speed}
\end{thm}
\bigskip

\begin{proof}
The proof is based on Kingman's subadditive theorem (theorem 2.4). For any probability $\P_{\rho_{inv}}$ generated by a minimal invariant state, the dynamical system $(\Omega,\P_{\rho_{inv}},\theta)$ is ergodic (see \cite[Theorem 1.1]{benoist2018entropy}). Set $q_n^{p,i}(\omega):=\Tr (M_i^p K_{\omega_n}\circ...\circ K_{\omega_0}(\hat\Xi_0))$ and $g_n^{p,i}(\omega):=\log(q_n^{p,i}(\omega))$.
Due to submultiplicativity of trace norm, the sequence $(g_n^{p,i})_{n\geq0}$ is almost subadditive:
$$g_{n+m}^{p,i}(\omega) \leq g_n^{p,i}(\omega) +g_m^{p,i}(\theta^n(\omega))-\log(\lambda^{p}_i)$$
with $\lambda^{p}_i$ such that $M^p_i \hat\Xi_0 \leq \lambda^{p}_i M^p_i$.
We first consider the behaviour of $g_n^{p,i}$ under a probability $\P_{j}^{\p}$.
Using Kingman's theorem, $\frac{1}{n}g_n^{p,i}$ converges $\P_{j}^{\p}$-a.s to $-S(\P_j^{\p},\P_i^p) -S(\P_j^{\p})$. Then $\frac{1}{n}\log \left(\frac{q_n^{p,i}}{q_n^{\p,j}}\right)$ converges $\P_j^{\p}$ -a.s to $-S(\P_j^{\p},\P_i^p)$. Under (ID), this last quantity is strictly negative (see \cite[Proposition 2.2]{benoist2018entropy}).

To get back to the probability $\P_{\rho_0}^{\p}$, we make use of a result stated in \cite{benoist2022selection}: under (ID), for any shift-invariant event A, we have
$$\P_{\rho_0}^{\p}(A)=\sum_{j}\tr(M_j^{\p}\rho_0)\P_{j}^{\p}(A).$$
In addition:
$$\P_{\rho_0}^{\p}(A)=\sum_{j}\tr(M_j^{\p}\rho_0)\P_{j}^{\p}(A~|~\I=j),$$
where $\I$ is the selection random variable appearing in theorem (2.1).
Identifying the two expressions above allows to announce that $\frac{1}{n}\log\o\frac{q_n^{p,i}}{q_n^{\p,\I}}\c$ converges to $S(\P_{\I}^{\p},\P_i^p)$ , $\P_{\rho_0}^{\p}$-a.s.
Finally,
\begin{align*}
\lim\limits_{n \to \infty}\frac{1}{n}\log\left(\frac{\pi_n^p}{\pi_n^{\p}}\right)&=\lim\limits_{n \to \infty}\frac{1}{n}\log\left(\frac{\sum_{i}q_n^{p,i}}{q_n^{\p,\I}}\right)\\
&\leq ~\underset{i}{max}\left( \lim\limits_{n \to \infty}\frac{1}{n}\log\left(\frac{q_n^{p,i}}{q_n^{\p,\I}}\right)\right)\\
&\leq ~\underset{i,j}{max}\left( \lim\limits_{n \to \infty}\frac{1}{n}\log\left(\frac{q_n^{p,i}}{q_n^{\p,j}}\right)\right)\\
&= ~\underset{i,j}{max} (-S(\P_{j}^{\p},\P_i^p)).
\end{align*}
Thus
$$ \lim\limits_{n \to \infty}\frac{1}{n} \log \left( \frac{ \pi_{n}^p}{ \pi_{n}^{\p}} \right) \leq -~\underset{i,j}{min} (S(\P_{j}^{\p},\P_i^p)),$$ which finishes the proof.
\end{proof}

\addtolength{\textheight}{-3cm}   % This command serves to balance the column lengths
                                  % on the last page of the document manually. It shortens
                                  % the textheight of the last page by a suitable amount.
                                  % This command does not take effect until the next page
                                  % so it should come on the page before the last. Make
                                  % sure that you do not shorten the textheight too much.

%%%%%%%%%%%%%%%%%%%%%%%%%%%%%%%%%%%%%%%%%%%%%%%%%%%%%%%%%%%%%%%%%%%%%%%%%%%%%%%%

\section{An algorithm for a parameter in a continuous set}
\label{sec:algo}
\subsection{Description of the algorithm}
Suppose that no \textit{a priori} information exists on the value of the parameter and that $\mathcal P$ is a continuous set. What would happen if we run trajectory $(\hat\Xi_n)$ and neither $a$ nor $b$ is equal to the true value $\p$? Repeating the same arguments used in the proof of theorem 4.1, it can be shown that $(\hat\Xi_n)$ still selects one of the parameters $a$ or $b$.\\

Assume first quantum channels $\Phi^a$, $\Phi^b$ and $\Phi^{\p}$ each have a unique invariant state. Denote by $\P^a$, $\P^b$ and $\P^{\p}$ the respective generated probability measures. With Kingman's theorem again, we can establish that $\frac{1}{n}\log(\frac{\pi_n^a}{\pi_n^b})$ converges to $S(\P^{\p},\P^b)-S(\P^{\p},\P^a)$. If this last quantity is negative, i.e. $S(\P^{\p},\P^b)<S(\P^{\p},\P^a)$, then $\pi_n^a \to 0$ and $\pi_n ^b \to 1$, and vice versa if $S(\P^{\p},\P^a)<S(\P^{\p},\P^b)$.
In brief, $(\hat\Xi_n)$ will select the parameter which is the closest to $\p$ with respect to the distance $\mathds d(\p,p):=S(\P^{\p},\P^{p})$.
In the same vein, when the quantum channels have more than one minimal invariant state, $(\hat\Xi_n)$ selects the closest parameter with respect to the distance $\mathds d(\p,p)=\underset{i}{min} (S(\P_\I^{\p},\P_i^{p}))$.\\

Based on these considerations, we can construct an algorithm to approach $\p$ by testing two values at a time and refining more and more the procedure.\\
Assuming that there is a set of measurement data $(y_0,y_1,...,y_T)$, and that $\p$ is a real number belonging to an interval $[u,v]$.\\

\noindent1. Initialize $a= u+\frac{1}{3}(v-u)$ ; $b= u+\frac{2}{3}(v-u)$ . Set a threshold $\epsilon>0$.\\
2. Run trajectory $(\hat\Xi_n)$ with data $(y_0,y_1,...,y_T)$ until $\pi_n^a$ or $\pi_n^b >$ 1- $\epsilon$\\
3. Take $s\in \{a,b\}$ the selected value such that $\pi_n^s>$ $1-\epsilon$ and redefine $a,b= s-\frac{1}{3}(b-a) ~,~ s+\frac{1}{3}(b-a)$    \\
4. Go to step 2\\

After a while, the algorithm may stop at step 2 when a and b become very close: neither $\pi_T^a$ nor $\pi_T^b$ is larger than $1-\epsilon$. All available data will have been exhausted; the last selected value constitutes an approximation of $\p$.

\subsection{Simulations results}
Let us take $\H=\mathbb C^2$ and two Kraus operators
$K_0(.)=V_0~.~V_0^\dag$ and $K_1(.)=V_1~.~V_1^\dag$ with 
$V_0=\begin{pmatrix} \frac{\sqrt{p}}{\sqrt{3}} & 0 \\
0&\frac{1}{2}\end{pmatrix}$,  $V_1=\begin{pmatrix} 0 & \frac{\sqrt{3}}{2} \\ 
\frac{\sqrt{3-p}}{\sqrt{3}}&0\end{pmatrix}$, and $p\in [0,3]$.\\

To generate a set of measurements records $(y_0,y_1,...,y_{T})$, we simulate the trajectory $(\rho_n)$ with $p=\p$ until a time T=2000, and then use this set of records to run the algorithm. Figure 1 presents simulation results until the algorithm is no longer able to discriminate between the two values.

An interesting phenomenon occurs on the 3rd run. Whereas the tested parameter on the right is closer to $\p$ than the tested parameter on the left with respect to the euclidian distance, the left parameter is selected, which shows that it is closer with respect to the distance $\mathds d$. Despite this kind of phenomenon, if $p\mapsto \mathds d(\p,p)$ is convex (or even simply decreasing on the left side of $\p$ and increasing on the right side), the algorithm will still come back to the true parameter at the end.

A generalization of this algorithm for a parameter in a k-dimensional space is possible. We then no longer test 2 values at a time, but rather $2^k$ values distributed in the parameter space and repeat the procedure around the selected one.
\begin{center}
\begin{figure}[htp]
%\centering
\includegraphics[scale=0.5]{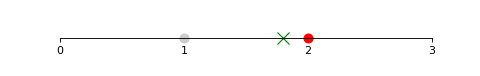}
\includegraphics[scale=0.5]{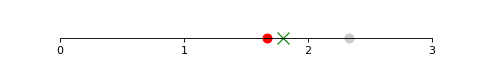}
\includegraphics[scale=0.5]{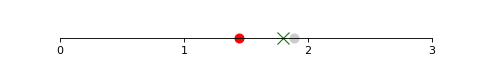}
\includegraphics[scale=0.5]{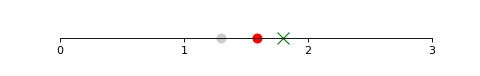}
\includegraphics[scale=0.5]{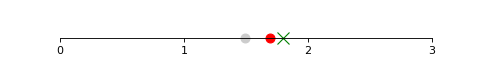}
\includegraphics[scale=0.5]{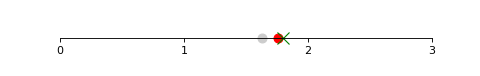}
\includegraphics[scale=0.5]{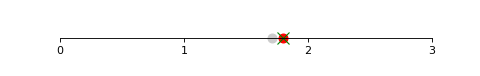}
\includegraphics[scale=0.5]{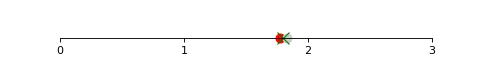}
\caption{Estimation of $\p=1.8$ marked with a green cross. Tested values marked with circles, selected value colored in red.}
\end{figure}
\end{center}
\section{Conclusion and further work}
\label{sec:conclusion}
In this paper, we show the convergence of the method developed in \cite{six2015parameter} to decide among several parameters which one is actually at stake in the physical setting. This method does not require the knowledge of the initial state of the quantum trajectory; it is only based on the measurement results obtained over time. We find a strictly negative Lyapunov exponent for the ratio of interest $\frac{\pi_n^p}{\pi_n^{\p}}$, showing the method is exponentially efficient. When the parameter belongs to an infinite set, we provide an algorithm to approach its true value.
In future work, we will further investigate the latter case and develop a rigorous estimator.

%%%%%%%%%%%%%
\section{Acknowledgments}The authors thank P. Rouchon for interesting and motivating discussions on parameter estimation for quantum trajectories. M. Bompais and C. Pellegrini thank T. Benoist for the ideas around the ergodic theory of quantum trajectories. This work is supported by the Agence Nationale de la Recherche projects Q-COAST ANR- 19-CE48-0003, QUACO ANR-17-CE40-0007, and IGNITION ANR-21-CE47-0015. 
C.P.\ has been supported by the ANR projects ``Quantum Trajectories'' ANR- 20-CE40-0024-01 and ``Evolutions Stochastiques Quantiques'' ESQuisses Projet-ANR-20-CE47-0014 and ``Investissements d'Avenir'' ANR-11-LABX-0040 of the French National Research Agency (ANR).

%The authors gratefully acknowledge the contribution of National Research Organization and reviewers' comments.

%%%%%%%%%%%%%%%%%%%%%%%%%%%%%%%%%%%%%%%%%%%%%%%%%%%%%%%%%%%%%%%%%%%%%%%%%%%%%%%%

\bibliographystyle{unsrt}
\bibliography{refs.bib}

\end{document}